\documentclass[a4paper,12pt]{article}
\usepackage{amsmath}
\usepackage{amssymb}
\usepackage{graphicx}
\usepackage{setspace}
\usepackage{multirow}
\begin{document}
\title{\textbf{Application of Cross-correlation Greens Function along with FDTD for Fast Computation of Envelope Correlation Coefficient over Wideband for MIMO Antennas}}

\author{\small Debdeep Sarkar and Kumar Vaibhav Srivastava \\
        \small Department of Electrical Engineering \\
        \small Indian Institute of Technology Kanpur, Kanpur, Uttar Pradesh, 208016, India \\
\small E-mail: debdeep1989@gmail.com and kvs@iitk.ac.in}
\date {}
\maketitle

\begin{abstract}
In this work\footnote{Extended version of this work has been accepted for publication in IEEE Transactions on Antennas and Propagation. \\ \textbf{IEEE Copyright Notice:}   
Copyright \copyright 2016 IEEE \\
Personal use of this material is permitted. Permission from IEEE must be obtained for all other
uses, in any current or future media, including reprinting/republishing this material for advertising or promotional purposes, creating new collective works, for resale or redistribution to servers or lists, or reuse of any copyrighted component of this work in other works.}, the concept of cross-correlation Green's functions (CGF) is used in conjunction with the finite difference time domain (FDTD) technique for calculation of envelope correlation coefficient (ECC) of any arbitrary MIMO antenna system over wide frequency band. Both frequency-domain (FD) and time-domain (TD) post-processing techniques are proposed for possible application with this FDTD-CGF scheme. The FDTD-CGF time-domain (FDTD-CGF-TD) scheme utilizes time-domain signal processing methods and exhibits significant reduction in ECC computation time as compared to the FDTD-CGF frequency domain (FDTD-CGF-FD) scheme, for high frequency-resolution requirements. The proposed FDTD-CGF based schemes can be applied for accurate and fast prediction of wideband ECC response, instead of the conventional scattering parameter based techniques which have several limitations. Numerical examples of the proposed FDTD-CGF techniques are provided for two-element MIMO systems involving thin-wire half-wavelength dipoles in parallel side-by-side as well as orthogonal arrangements. The results obtained from the FDTD-CGF techniques are compared with results from commercial electromagnetic solver Ansys HFSS, to verify the validity of proposed approach. 
\end{abstract}




\section{Introduction}
Multiple-Input Multiple-Output (MIMO) antenna systems are the key elements for the current 4G and future 5G wireless networks, which have the potential to support high data rates of more than 1 Gbps \cite{mimo1}-\cite{mimo2}. Among the different performance metrics for MIMO antennas, envelope correlation coefficient (ECC) is one of the most important parameter as it characterizes the mutual correlation between the communication channels in any multi-antenna system \cite{mimo3}. Different diversity techniques like spatial diversity, pattern diversity and polarization diversity are employed to ensure low value of ECC in the working bandwidth of the MIMO antenna system \cite{mimo4}-\cite{mimo6}. Thus, characterizing the ECC for a given MIMO configuration over a specified antenna bandwidth is a problem of utmost importance for system. The general formula for computation of ECC utilizes the radiated far-fields of the individual antennas, with other antennas terminated in matched load condition \cite{mimo3}-\cite{ecc1}. But this procedure requires complete evaluation of the 3D radiation patterns at all the frequencies in the antenna bandwidth. This makes the process tedious in terms of measurement as well as computation. In \cite{ecc2}, it is proved from a simple derivation that ECC can be determined using the S-parameters of the multi-antenna system, when lossless and single-mode antennas are considered. This S-parameter based evaluation of ECC considerably reduces the difficulties of the far-field based method, but it suffers from some serious drawbacks. It is pointed out in \cite{ecc3} that when the antenna system is lossy, which is the case in any practical situation, the S-parameter based formula as derived in \cite{ecc2} is not valid. Some corrective measures for ECC calculation of lossy antenna systems are suggested in \cite{ecc3}, which involve inclusion of the antenna efficiencies in the S-parameter based expression of ECC. However it is well pointed out in \cite{ecc4}-\cite{antar1}, that a lot of issues still exist with this S-parameter based ECC calculation. It is elaborated in \cite{antar1} that the S-parameter based method can give erroneous results for cases with zero or weak mutual coupling between antenna elements. Also when strong mutual coupling is present between the two excited antennas, and an electromagnetically coupled third object (scatterer or another antenna) is also present in the surrounding environment, the S-parameters based formula becomes ambiguous and produces wrong results \cite{antar1}. It is commented in \cite{antar1} that ECC of a MIMO system depends on the entire current on the antenna surfaces. This implies that it is not possible to compute ECC based on S-parameters only, since S-parameters are relevant only to the currents at the physical antenna ports. \\

A useful alternative technique for ECC calculation is recently introduced by researchers, which uses spatial antenna current distributions instead of the 3D far-field patterns of the antennas \cite{antar1}-\cite{antar4}. This technique requires the evaluation of Cross-correlation Green's functions (CGF), along with the amplitude and phase distribution of antenna currents on the antenna structure, to accurately determine the ECC \cite{antar1}-\cite{antar4}. The effectiveness of this CGF-based procedure relies on approximating the current distribution on any given antenna structure into a combination of infinitesimal radiating dipole elements having currents of appropriate amplitude and phase. The effects of all such infintesimal dipole elements are combined using CGF to obtain the desired ECC value \cite{antar1}. A Method-of-Moments based approach using CGF is shown in \cite{antar1}, for calculating ECC and diversity gain of thin-wire dipole antenna systems. Also genetic algorithm (GA) based techniques are formulated in \cite{antar2}-\cite{antar3}, to predict the optimum antenna current distributions that can provide the best ECC and diversity performance. However, the procedures introduced in \cite{antar1}-\cite{antar4} are mainly targeted for evaluation of ECC of MIMO antenna systems at a single frequency. Since recent focus of MIMO antenna designers is on development of multi-band and ultrawideband MIMO antennas \cite{mant1}-\cite{mant3}, fast ECC evaluation over a wide frequency band is desired for practical purposes. The S-parameter based method, although flawed in several respects, facilitates this fast and wideband computation of ECC for a given MIMO antenna system. This is why many antenna designers still resort to this S-parameter based method for ECC calculation in multi-band or wideband MIMO antenna systems, inspite of the ambiguities present in it \cite{ecc4}-\cite{ecc5}.  \\

In this work, the CGF-based procedure of ECC calculation of \cite{antar1}-\cite{antar4} is integrated with finite difference time domain (FDTD) simulation to obtain wideband frequency variation of ECC for a given MIMO antenna system. This technique is designated as FDTD-CGF scheme throughout this manuscript. Although FDTD simulations have been applied in MIMO system analysis \cite{fdtd1}-\cite{fdtd22}, integration of CGF approach with FDTD for ECC computation has not been previously reported, to the best of the knowledge of the authors. Two different post-processing techniques for this FDTD-CGF based scheme are proposed, one based on frequency-domain (FD) calculations, another using time-domain (TD) methods. The TD-based post-processing gives wideband ECC results in very less computation time and good accuracy, which makes it more suitable as compared to the FD-scheme. Although FD post-processing of CGF using MoM based antenna analysis is reported in \cite{antar1}, such TD post-processing method of CGF has not been reported in literature, to the best of the knowledge of the authors. At this point, it should be noted that the prescribed FDTD-CGF technique is fundamentally different from the discrete Green's function (DGF) based FDTD formulation as discussed in \cite{dgf1}-\cite{dgf5}. The purpose of DGF-FDTD is mainly to use FDTD from a system point of view, such that the accuracy and efficiency of the FDTD method is improved \cite{dgf2}. Applications of DGF-FDTD for simulations of wire antennas that are mutually coupled with inhomogeneous dielectric objects as well as for UWB antenna analysis have been reported \cite{dgf3}-\cite{dgf4}. However, DGF-FDTD has not been directly applied for MIMO antenna systems. The proposed FDTD-CGF technique utilizes conventional Yee's FDTD grids and continuous CGFs for post-processing and ECC computation, unlike the DGF-FDTD technique. Although there is possibility of using CGFs along with DGF-FDTD, that aspect is not explored in this paper. \\

The entire paper is organized as follows. Detailed formulations of the FDTD-CGF-FD and FDTD-CGF-TD techniques are provided in section \ref{s2}. In section \ref{s3}, effectiveness of these FDTD-CGF based procedures are validated by use of some numerical examples. Two MIMO antenna configurations (parallel side-by-side and orthogonal) of thin-wire half-wavelength dipoles are considered for analysis. The results of the in-house FDTD codes are also compared with that obtained from commercially available FEM-based Ansys HFSS, to illustrate the correctness of the proposed computation techniques. Section \ref{s4} includes some discussions on the aspects of simulation time and relative accuracy for the FDTD-CGF-TD technique as compared to FDTD-CGF-FD. Finally some concluding remarks are provided in section \ref{s5}.   
       
\section{ECC Computation using FDTD-CGF Scheme} \label{s2}
This section provides the stepwise formulation of the FDTD-CGF technique for calculation of ECC of any MIMO antenna system dover wide frequency range. From the definition as discussed in \cite{mimo3},\cite{ecc1}, the ECC (or $\rho_{e,12}$) between two antennas (designated as antenna-1 and antenna-2) can be obtained from the 3D far-field radiation patterns of the individual antennas from the formula: 
\begin{equation} \label{ecc}
\rho_{e,12}(\omega)=\frac{|\int_{4\pi} d\Omega \; \mathbf{E}_{1}(\theta,\phi,\omega) \cdot \mathbf{E}^{*}_{2}(\theta,\phi,\omega)|^{2}}{{\int_{4\pi} d\Omega \; |\mathbf{E}_{1}(\theta,\phi,\omega)|^{2}}{\int_{4\pi} d\Omega \; |\mathbf{E}_{2}(\theta,\phi,\omega)|^{2}} }
\end{equation}

It is assumed that the antennas 1 and 2 are radiating in uniform multipath environment of balanced polarization (i.e., an isotropic environment) \cite{ecc1}. The radiation patterns due to antennas 1 and 2 operating at angular frequency $\omega$ are denoted by $\mathbf{E}_{1}(\theta,\phi,\omega)$ and $\mathbf{E}_{2}(\theta,\phi,\omega)$ respectively. The integration with respect to the full solid angle $\Omega$ is represented by $\int_{4\pi}$. The spatial current-density distributions when antennas 1 and 2 are excited respectively at angular frequency $\omega$ are represented by $\mathbf{J_{1}}(\mathbf{r},\omega)$ and $\mathbf{J_{2}}(\mathbf{r},\omega)$ respectively. The field cross-correlation term in the numerator expression of $\rho_{e,12}(\omega)$ as shown in \eqref{ecc}, can be evaluated from the antenna current distributions using the following formula \cite{antar1}:
\begin{multline} \label{rhonum}
\int_{4\pi} d\Omega \; \mathbf{E}_{1}(\theta,\phi,\omega) \cdot \mathbf{E}^{*}_{2}(\theta,\phi,\omega) = \int_{V_{1}} d^{3}r' \int_{V_{2}} d^{3}r'' \; \mathbf{J_{1}}(\mathbf{r'},\omega) \cdot \mathbf{\bar{C}}(\mathbf{r'},\mathbf{r''},\omega) \cdot \mathbf{J^{*}_{2}}(\mathbf{r''},\omega) \\ =\int_{V_{1}} d^{3}r' \int_{V_{2}} d^{3}r'' \sum_{p}\sum_{q} \Psi_{12pq}(\mathbf{r'},\mathbf{r''},\omega)
\end{multline}    

Here $\mathbf{\bar{C}}(\mathbf{r'},\mathbf{r''},\omega)$ is the dyadic tensor representation for the cross-correlation Green's function (CGF) as defined in \cite{antar1}. The symbol ``$*$'' denotes the complex conjugate operation. It is assumed the two current-densities $\mathbf{J_{1}}(\mathbf{r},\omega)$ and $\mathbf{J_{2}}(\mathbf{r},\omega)$ are nonzero in the volume $V_{1,2}$, which are bounded subregions of the entire space $\mathbb{R}^{3}$ \cite{antar1}. The term $\Psi_{12pq}(\mathbf{r'},\mathbf{r''},\omega)$ where ($p=x,y,z$ and $q=x,y,z$), can be calculated from the components of antenna current-densities and dyadic CGF as follows:
\begin{equation} \label{psipq}
\Psi_{12pq}(\mathbf{r'},\mathbf{r''},\omega)=J_{1p}(\mathbf{r'},\omega)C_{pq}(\mathbf{r'},\mathbf{r''},\omega)J^{*}_{2q}(\mathbf{r''},\omega)
\end{equation}
Here $\mathbf{J}_{m}(\mathbf{r'},\omega) = J_{mx}(\mathbf{r'},\omega)\hat{x}+J_{my}(\mathbf{r'},\omega)\hat{y}+J_{mz}(\mathbf{r'},\omega)\hat{z}$ where $m=1,2$. The term $C_{pq}(\mathbf{r'},\mathbf{r''},\omega)$ can be expanded in a general form as:
\begin{equation} \label{cpq} 
C_{pq}(\mathbf{r'},\mathbf{r''},\omega)=\int_{0}^{\pi} \int_{0}^{2\pi} d\theta d\phi \; f_{pq}(\theta,\phi) e^{jk(\mathbf{r'}-\mathbf{r''}).\hat{r}} = \int_{0}^{\pi} \int_{0}^{2\pi} d\theta d\phi f_{pq}(\theta,\phi) e^{j\omega t_{d}(\theta,\phi)} 
\end{equation} 
Here $k=\omega/c$ denotes the wave-number, where $c$ is the velocity of electromagnetic wave in free space. The radial unit vector in spherical co-ordinate systems is designated by $\hat{r}=\hat{x}\sin\theta\cos\phi+\hat{y}\sin\theta\sin\phi+\hat{z}\cos\theta$. The function $t_{d}(\theta,\phi)$ standing for the $(\theta,\phi)$ dependent time-advancement (or time-delay), can be calculated as:
\begin{equation}
t_{d}(\theta,\phi)=\frac{1}{c}(x_{d}\sin\theta\cos\phi + y_{d}\sin\theta\sin\phi + z_{d}\cos\theta)
\end{equation}
Using $\mathbf{r'}=x'\hat{x}+y'\hat{y}+z'\hat{z}$ and $\mathbf{r''}=x''\hat{x}+y''\hat{y}+z''\hat{z}$, $x_{d},y_{d}$ and $z_{d}$ can be evaluated as $(x'-x'')$, $(y'-y'')$ and $(z'-z'')$ respectively. The expressions for the angular cross-correlation component functions $f_{pq}(\theta,\phi)$ are obtained from \cite{antar1} and provided in the appendix for the sake of completeness. Substituting  expression of $C_{pq}$ from \eqref{cpq} into the expression of $\Psi_{12pq}(\mathbf{r'},\mathbf{r''},\omega)$, one can write: 
\begin{multline} \label{psipq2}
\Psi_{12pq}(\mathbf{r'},\mathbf{r''},\omega)=J_{1p}(\mathbf{r'},\omega)C_{pq}(\mathbf{r'},\mathbf{r''},\omega)J_{2q}(\mathbf{r''},\omega) \\ =\int_{0}^{\pi} \int_{0}^{2\pi} d\theta d\phi \; f_{pq}(\theta,\phi) \left[ e^{j\omega t_{d}(\theta,\phi)} J_{1p}(\mathbf{r'},\omega) J^{*}_{2q}(\mathbf{r''},\omega) \right]
\end{multline}
Taking Inverse Fourier transform (IFT) of the expressions in both sides of \eqref{psipq2} and utlizing Fourier transform properties like time-delay, complex conjugation and convolution \cite{fourier}, we can express the time domain function $\psi_{12pq}(\mathbf{r'},\mathbf{r''},t)$ as:
\begin{equation} \label{psipqt}
\psi_{12pq}(\mathbf{r'},\mathbf{r''},t) = \int_{0}^{\pi} \int_{0}^{2\pi} d\theta d\phi \; f_{pq}(\theta,\phi) \xi_{12pq}(\mathbf{r'},\mathbf{r''},t+t_{d}(\theta,\phi))
\end{equation} 

where the term $\xi_{12pq}(t)$ is be denoted by:
\begin{equation} \label{psipqt2}
\xi_{12pq}(\mathbf{r'},\mathbf{r''},t)=J_{1p}(\mathbf{r'},t) \oplus J_{2q}(\mathbf{r''},-t)
\end{equation} 
The symbol ``$\oplus$'' stands for time-domain convolution. The $-t$ present in the argument of $J_{2q}$ in \eqref{psipqt2} indicates time-reversal operation on the vector $J_{2q}(\mathbf{r''},t)$. It can be observed that Fourier transform of $\psi_{12pq}(\mathbf{r'},\mathbf{r''},t)$ as in \eqref{psipqt} yields the LHS term $\Psi_{12pq}(\mathbf{r'},\mathbf{r''},\omega)$ as shown in \eqref{psipq}. Since Fourier transform is a linear operation, one can use Fourier transform of \eqref{psipqt} in \eqref{rhonum}. Interchanging the order of summation and Fourier transform operation to evaluate the field cross-correlation term as:
\begin{equation} \label{rhonum2}
\int_{4\pi} d\Omega \; \mathbf{E}_{1}(\theta,\phi,\omega) \cdot \mathbf{E}^{*}_{2}(\theta,\phi,\omega)= \mathcal{F} \left[ \int_{V_{1}} d^{3}r' \int_{V_{2}} d^{3}r'' \sum_{p}\sum_{q} \psi_{12pq}(\mathbf{r'},\mathbf{r''},t) \right]
\end{equation}
Here the symbol ``$\mathcal{F}$'' denotes the Fourier transform operation. Similarly, the field self-correlation terms present in the denominator expression of $\rho_{e,12}(\omega)$ as shown in \eqref{ecc} can be written as (for $m=1,2$):
\begin{multline} \label{rhoden}
\int_{4\pi} d\Omega \; |\mathbf{E}_{m}(\theta,\phi,\omega)|^{2} = \int_{4\pi} d\Omega \; \mathbf{E}_{m}(\theta,\phi,\omega) \cdot \mathbf{E}^{*}_{m}(\theta,\phi,\omega) \\ = \int_{V_{1}} d^{3}r' \int_{V_{2}} d^{3}r'' \; \mathbf{J_{m}}(\mathbf{r'},\omega) \cdot \mathbf{\bar{C}}(\mathbf{r'},\mathbf{r''},\omega) \cdot \mathbf{J^{*}_{m}}(\mathbf{r''},\omega) \\ = \int_{V_{1}} d^{3}r' \int_{V_{2}} d^{3}r'' \sum_{p}\sum_{q} \Psi_{mmpq}(\mathbf{r'},\mathbf{r''},\omega) 
\end{multline}

Furthermore, using the Fourier transform relation:
\begin{equation} \label{rhoden2}
\int_{V_{1}} d^{3}r' \int_{V_{2}} d^{3}r'' \sum_{p}\sum_{q} \Psi_{mmpq}(\mathbf{r'},\mathbf{r''},\omega) = \mathcal{F} \left[ \int_{V_{1}} d^{3}r' \int_{V_{2}} d^{3}r'' \sum_{p}\sum_{q} \psi_{mmpq}(\mathbf{r'},\mathbf{r''},t) \right]
\end{equation}

In the above expression, the term $\psi_{mmpq}(\mathbf{r'},\mathbf{r''},t)$ where $m=1,2$, can be calculated as:
\begin{equation} \label{psimmpqt}
\psi_{mmpq}(\mathbf{r'},\mathbf{r''},t) = \int_{0}^{\pi} \int_{0}^{2\pi} d\theta d\phi \; f_{pq}(\theta,\phi) \xi_{mmpq}(\mathbf{r'},\mathbf{r''},t+t_{d}(\theta,\phi))
\end{equation} 

The quantity $\xi_{mmpq}(t)$ where $m=1,2$, can be written as:
\begin{equation} \label{ximmpqt}
\xi_{mmpq}(\mathbf{r'},\mathbf{r''},t)=J_{mp}(\mathbf{r'},t) \oplus J_{mq}(\mathbf{r''},-t)
\end{equation}

In a typical FDTD scheme, the antennas are excited by impulse-like functions in time domain (like Gaussian or Differentiated Gaussian pulses). Hence desired information (mainly scattering parameters) over a broadband frequency range can be extracted by computing suitable Fourier transforms of time domain currents/voltages. It is quite evident here that if the antenna current-density distributions in time domain (i.e. $\mathbf{J_{1}}(\mathbf{r},t)$ and $\mathbf{J_{2}}(\mathbf{r},t)$) can be calculated from an FDTD simulation, the ECC for a broad frequency range denoted by $\rho_{e,12}(\omega)$ in \eqref{ecc}, can calculated by following two distinct methods as described below.  

\subsection{Method-1 (FDTD-CGF-FD)}
This method involves the calculation of the CGF using \eqref{rhonum} in frequency-domain (FD) itself to determine $\rho_{e,12}(\omega)$. The complex antenna current densities at different locations for different  $\omega$ are computed via Fourier transform of the antenna currents at those locations, and used alongwith the CGF in frequency domain. The steps are as follows:
\begin{enumerate}
\item The component of CGF $C_{pq}(\mathbf{r'},\mathbf{r''},\omega)$ for each $\omega$ is calculated using \eqref{cpq}. 
\item By taking Fourier transforms of $\mathbf{J_{1}}(\mathbf{r},t)$ and $\mathbf{J_{2}}(\mathbf{r},t)$, the frequency domain expressions of $\mathbf{J_{1}}(\mathbf{r'},\omega)$ and $\mathbf{J^{*}_{2}}(\mathbf{r''},\omega)$ as required to calculate $\Psi_{12pq}(\mathbf{r'},\mathbf{r''},\omega)$ in \eqref{psipq} are obtained. 
\item Using the obtained values of $\Psi_{12pq}(\mathbf{r'},\mathbf{r''},\omega)$ in  \eqref{rhonum}, the field cross-correlation term needed for the numerator expression of $\rho_{e,12}(\omega)$ is evaluated. 
\item To calculate the field self-correlation terms of \eqref{ecc}, $\Psi_{mmpq}(\mathbf{r'},\mathbf{r''},\omega)$ are determined from \eqref{rhoden} for each angular frequency $\omega$ in a similar fashion of step-2. 
\item Finally using \eqref{ecc}, the ECC for wide frequency range is computed using the values of field cross-correlation and self-correlation terms.    
\end{enumerate} 

\subsection{Method-2 (FDTD-CGF-TD)}
In this technique, instead of calculating the CGF directly in frequency-domain, its effect is embedded in the time-domain (TD) calculations by suitable signal-processing methods. Operations like time-reversal, convolution and integrations are performed on the temporal current density profiles, to get the desired $\rho_{e,12}(\omega)$. The following steps are involved: 
\begin{enumerate}
\item Once the values of $\mathbf{J_{1}}(\mathbf{r},t)$ and $\mathbf{J_{2}}(\mathbf{r},t)$ are obtained from FDTD simulations, the term $\xi_{pq}(\mathbf{r'},\mathbf{r''},t)$ is computed using \eqref{psipqt2}. 
\item Using suitable $f_{pq}(\theta,\phi)$ and performing numerical integration on its product with delayed/advanced versions of $\xi_{12pq}(\mathbf{r'},\mathbf{r''},t)$ as indicated in \eqref{psipqt}, the term $\psi_{12pq}(\mathbf{r'},\mathbf{r''},t)$ is evaluated. 
\item Once $\psi_{12pq}(\mathbf{r'},\mathbf{r''},t)$ is obtained, the field cross-correlation term can be calculated by summation, numerical integration and Fourier transform using \eqref{rhonum2}.
\item In a way similar to step-2 and step-3, \eqref{ximmpqt} and \eqref{psimmpqt} are used to calculate the terms $\psi_{mmpq}(\mathbf{r'},\mathbf{r''},t)$ for $m=1,2$.
\item After determining $\psi_{mmpq}(\mathbf{r'},\mathbf{r''},t)$ for $m=1,2$, the field self-correlation terms are obtained utilizing \eqref{rhoden2}. 
\item Once the numerator and denominator expressions for $\rho_{e,12}(\omega)$ are evaluated, broadband calculation of ECC is done in a straightforward fashion using \eqref{ecc}. 
\end{enumerate}

\vspace{10pt}

At this point some general comments can be made regarding the two above-mentioned post-processing methods (FDTD-CGF-FD and FDTD-CGF-TD) for computing $\rho_{e,12}(\omega)$. In the FDTD-CGF-FD technique, $\mathbf{J_{1}}(\mathbf{r'},\omega)$ and $\mathbf{J^{*}_{2}}(\mathbf{r''},\omega)$ can be obtained from their respective time-domain waveforms by Fourier transform. But the values of $C_{pq}(\mathbf{r'},\mathbf{r''},\omega)$ have to be calculated for all the frequency values within the range of interest. As observed from \eqref{cpq}, this involves evaluation of integrals that can be performed with the help of in-built quadrature routines of MATLAB. This is the step that consumes most of the post-processing time in FDTD-CGF-FD technique, when high frequency resolution is required. It should be noted here that the integral in \eqref{cpq} can also be evaluated analytically, as indicated in \cite{antar1}. This can potentially lead to reduction in the processing time of FDTD- CGF-FD. However, analytical evaluation of \eqref{cpq} is not done in this paper and this aspect will be addressed in future works. On the other hand, for the FDTD-CGF-TD technique, the quantities $\psi_{11pq}(\mathbf{r'},\mathbf{r''},t)$, $\psi_{12pq}(\mathbf{r'},\mathbf{r''},t)$ and $\psi_{22pq}(\mathbf{r'},\mathbf{r''},t)$ can be determined directly in time-domain by signal processing operations on the FDTD-computed current waveforms like time-reversal, convolution, delay (or advancement) and integration as shown in \eqref{psipqt} and \eqref{psimmpqt}. These operations consume much less time, when performed upon time-domain MATLAB vectors. Clearly, the steps in the FDTD-CGF-TD technique are not directly dependent on the desired frequency resolution, since no requirement of calculating CGF for all the frequencies is there. It can be observed that the effects of CGF have been embedded in the evaluation of integrals in \eqref{psipqt2} and \eqref{ximmpqt}. These integrals can be further reduced to discrete double-summation expressions, by considering small incremental angles $\Delta\theta$ and $\Delta\phi$. This double summation approximation of the integrals in \eqref{psipqt} and \eqref{psimmpqt} significantly speeds up the process of calculating $\psi_{11pq}(\mathbf{r'},\mathbf{r''},t)$, $\psi_{12pq}(\mathbf{r'},\mathbf{r''},t)$ and $\psi_{22pq}(\mathbf{r'},\mathbf{r''},t)$. Finally, one needs to perform the Fourier transforms of these three terms only for computing the ECC value using \eqref{ecc}. However the accuracy of the evaluated ECC, depends on the values of the incremental angles $\Delta\theta$ and $\Delta\phi$, as will be evident from the discussion in section \ref{s4}. Thus can be intuitively concluded that although FDTD-CGF-FD is inherently more accurate for broadband evaluation of $\rho_{e,12}(\omega)$, the computation time will be increased considerably if the desired frequency resolution is very high. On the other hand, FDTD-CGF-TD technique will be comparatively faster to calculate $\rho_{e,12}(\omega)$, but the accuracy will largely depend upon the choice of incremental angles $\Delta\theta$ and $\Delta\phi$ for numerical evaluation of the integral in \eqref{psipqt} and \eqref{psimmpqt}. These intuitive predictions will be further elaborated upon in the next section, using the results obtained via FDTD simulation and subsequent post-processing.  

\section{Numerical Examples and Results} \label{s3}
In this section, ECC over a wide frequency range is calculated for two MIMO antenna systems, by using the two methods discussed in the previous section. The MIMO antenna systems under consideration consist of two thin-wire (wire radius $\ll \lambda_{0}$, where $\lambda_{0}$ is the operating wavelength) dipole antennas. Instead of using Method of Moments based formulation suggested in \cite{antar1}, conditionally-stable FDTD technique with uniform cartesian grid \cite{fdtd3},\cite{fdtdds} is used for modelling the antennas. The currents on every antenna segment as well as the port-voltages and currents are computed in time-domain. To validate the accuracy of the FDTD simulation setup, the same system is modelled and simulated in commercially available FEM based Ansys HFSS. It will be shown that the results from FDTD and HFSS simulations are overall in good agreement.

\begin{figure}[htbp]
\begin{center}
\includegraphics[width=13cm]{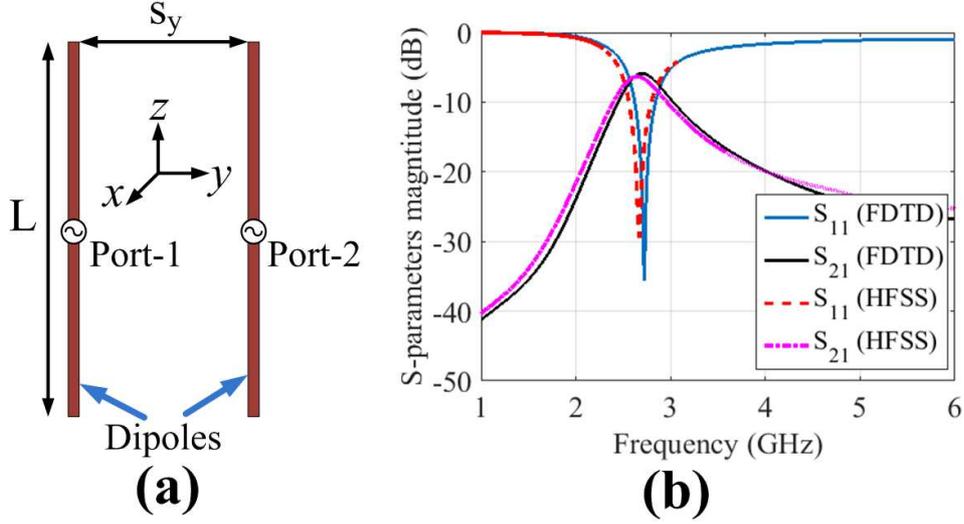}
\caption{\label{f1} (a) MIMO antenna system-1: Topology for the two thin-wire dipole antenna system arranged in side-by-side configuration in free space. Here $L=50$ mm, $S_{y}=20$ mm. (b) Comparative plots of S-parameters (magnitude) with frequency, as obtained from FDTD and HFSS simulations.}
\end{center}
\vspace{-10pt}
\end{figure} 

\subsection{Two Parallel Side-by-Side Dipoles}
Fig. \ref{f1} shows the first two element MIMO configuration comprising of two $z$-directed center-fed thin-wire dipole antennas operating at $f_{0}=2.70$ GHz. The desired frequency range of analysis is chosen as $1-6$ GHz with frequency resolution $10$ MHz, without the loss of any generality. The length $L$ for each dipole is kept at $0.45\lambda_{0}$, where $\lambda_{0}=c/f_{0}$. The dipoles are kept closely spaced in a side-by-side configuration, with the inter-element spacing at $S_{y}=0.18\lambda_{0}$, which is typical for such MIMO antenna systems. The FDTD grids in cartesian coordinates are chosen to be cubical having dimensions of  $\Delta x=\Delta y=\Delta z=\Delta=2.5$ mm, hence the cross-sections of the wire-dipoles are assumed to be square instead of cylindrical. Absorbing boundary condition is ensured by truncating the problem space with $10$ cells of Berenger's Perfectly matched layer (PML) in every surrounding direction. The entire problem space size is $60\Delta \times 60\Delta \times 60\Delta$. The time-step for FDTD marching procedure is kept at $\Delta t=3.833$ ps, satisfying the well-known CFL criterion \cite{taflove},\cite{fdtd3}. The total FDTD simulation time is taken sufficiently large ($3000\Delta t$) such that the initial transients get receded and steady-state condition is achieved. The thin-wire dipoles are assumed to be made of perfect electric conductors (PEC) and are modelled by forcing $z$-component of the electric field along the conductors of the dipole arms to be zero \cite{fdtd3}. One cell Delta-gap scheme \cite{fdtd3} is employed to excite each center-fed dipole at a time with a unit-amplitude differentiated Gaussian pulse $p(t)$ represented by:
\begin{equation} \label{gpt}
p(t)=-\frac{(t-\tau)}{\sigma}e^{\frac{-(t-\tau)^{2}}{2\sigma^{2}}}
\end{equation}     
The spread parameter $\sigma$ for $p(t)$ as shown in \eqref{gpt} is kept at $3\Delta t$ such that the desired frequency range of $1-6$ GHz is covered. The dipoles are excited to the peak value after a delay-time $\tau$ starting from the beginning of the simulation. Assuming the center location of the dipole-$m$ ($m=1,2$) under consideration to be $(i_{cm},j_{cm},k_{cm})$ in the FDTD space, the input-voltage for that dipole at $n$-th instant can be obtained as:
\begin{equation} \label{vin}
V_{in,m}^{n}=-E_{z}(i_{cm},j_{cm},k_{cm})\Delta z
\end{equation}

The thin-wire dipole antenna is considered as a linear arrangement of a number of infinitesemal current elements located at the FDTD grids. For calculation of the current at each of these elements, we use Ampere's circuital theorem \cite{fdtd3} to get the following expression:
\begin{multline} \label{Itot}
I_{m}^{n+\frac{1}{2}}(k) = \Delta x \left( H_{x}^{n+\frac{1}{2}}(i_{cm},j_{cm}-\frac{1}{2},k)-H_{x}^{n+\frac{1}{2}}(i_{cm},j_{cm}+\frac{1}{2},k) \right) \\ + \Delta y \left( H_{y}^{n+\frac{1}{2}}(i_{cm}+\frac{1}{2},j_{cm},k)-H_{y}^{n+\frac{1}{2}}(i_{cm}-\frac{1}{2},j_{cm},k) \right)
\end{multline}  
Here $k$ ranges from $k_{sm}+1/2$ to $k_{em}+1/2$, where the $E_{z}$ field is forced to be zero at all the locations starting from $(i_{cm},j_{cm},k_{sm})$ to $(i_{cm},j_{cm},k_{em})$, except at the feed-gap that is $(i_{cm},j_{cm},k_{cm})$ (Here $m=1,2$). For calculating the input port current $I_{in,m}^{n+\frac{1}{2}}$ of dipole-$m$, $k=k_{cm}+1/2$ is subsituted in \eqref{Itot}. The corresponding current-density values can be determined using \eqref{Itot} and the relation $J_{m}^{n+\frac{1}{2}}(k)=I_{m}^{n+\frac{1}{2}}(k)/\Delta x \Delta y$ (for $m=1,2$). Assuming that the dipoles are identical ($Z_{11}=Z_{22}$) and the two-dipole system to be reciprocal ($Z_{12}=Z_{21}$), the $Z$-parameters for the two-dipole antenna system in Fig. \ref{f1}(a) is determined. One dipole is excited at a time and the corresponding induced port-voltages as well port-currents are calculated, from which the $Z$-parameters are determined. Using relations from \cite{pozar}, the $Z$-parameters are then converted to the scattering ($S$) parameters for the system. In Fig. \ref{f1}(b), the $S_{11}$ and $S_{21}$ for the two dipole antenna system (Fig. \ref{f1}(a)) as simulated in FDTD as well as Ansys HFSS are shown in the desired frequency range of $1-6$ GHz. Strong mutual coupling between the two dipoles is observed from the $S_{21}$ response as shown in Fig. \ref{f1}(b). 

\begin{figure}[htbp]
\begin{center}
\includegraphics[width=9cm]{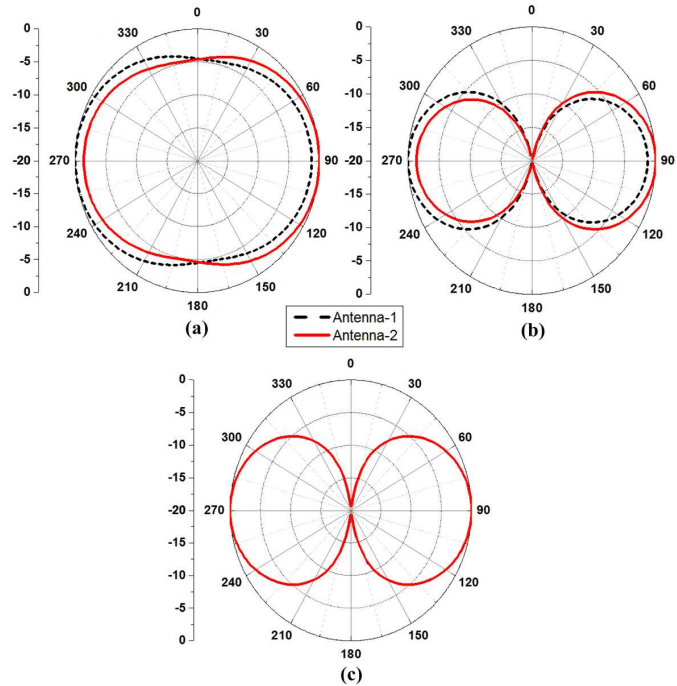}
\caption{\label{f2} Simulated total-gain patterns at resonance frequency, for both the antennas in the MIMO configuration of  Fig. \ref{f1}(a) along the three principal planes: (a) $xy$-plane, (b) $yz$-plane and (c) $xz$-plane.}
\end{center}
\vspace{-10pt}
\end{figure} 

\begin{figure}[htbp]
\begin{center}
\includegraphics[width=9cm]{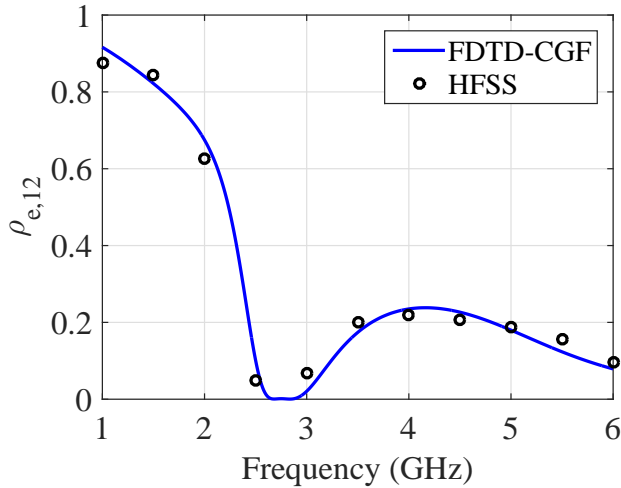}
\caption{\label{f3} Simulated variation of ECC for the MIMO configuration of Fig. \ref{f1}(a) over the frequency range $1-6$ GHz as obtained from FDTD-CGF techniques and Ansys HFSS. In Ansys HFSS, the ECC is calculated from far-field distribution at discrete frequency points.}
\end{center}
\vspace{-10pt}
\end{figure} 

To obtain the radiation patterns and ECC values, one needs to properly analyze the temporal current distributions on the infinitesimal elements constituting the dipoles in MIMO scenario. Thus when one dipole is excited by the Gaussian pulse, the other dipole has to be terminated by matched load. The technique suggested in \cite{fdtd3} is used to model this matched load termination as a $50\Omega$ resistor connected at the dipole port. When some excitation is provided at port-1, dipole-1 carries the main current whereas some current is also induced into dipole-2 due to the mutual coupling effects. Hence the total antenna current when port-1 is excited must account for the current in dipole-1 as well as the coupled current in dipole-2. In this way, the temporal profiles of the total antenna current densities ($\mathbf{J_{1}}(\mathbf{r},t)$ and $\mathbf{J_{2}}(\mathbf{r},t)$) for individual excitations of ports 1 and 2 are obtained. Next, fourier transform of the transient current densities at each of the dipole-cells are obtained. In this way, the amplitude and phase of current-densities are determined at the desired frequencies. By properly combining the effects of these complex current densities pertaining to the infinitesimal current elements through summation, one can calculate the radiation patterns for the MIMO configuration. Fig. \ref{f2} illustrates the total gain patterns when one antenna element is excited along the three principal planes ($xy$, $yz$ and $xz$). It can be observed that when one $z$-directed dipole in the MIMO configuration (Fig. \ref{f1}(a)) is excited, the other dipole acts as a strongly coupled parasitic reflector element, resulting in slightly directive pattern along the $xy$ and $yz$ planes. Along the $xz$ plane, the pattern due to individual antennas exactly overlap, as is evident from Fig. \ref{f1}(c). It should be noted that these radiation patterns shown in Fig. \ref{f2} are also verified by Ansys HFSS simulations. \\

ECC for the MIMO configuration of Fig. \ref{f1}(a) is calculated both via FDTD-CGF-FD and FDTD-CGF-TD techniques as discussed in the previous section. The frequency resolution considered is $0.01$ GHz, within the range of $1-6$ GHz. It is observed that, when the angular steps in computing the time-domain integrals of \eqref{psipqt} and \eqref{psimmpqt} are taken sufficiently small ($<0.1\pi$), both the methods (FDTD-CGF-FD and FDTD-CGF-TD) yield the same values of ECC, although the computation time needed for FDTD-CGF-TD technique is much lesser. This aspect of post-processing time will be highlighted in a later section. In order to validate the proposed FDTD-CGF based technique, the ECC is computed in Ansys HFSS using far-field based methods and plotted with frequency as shown in Fig. \ref{f3}. The ECC from HFSS has been computed for much lesser frequency resolution, since determination of the far-field patterns at each frequency point and subsequent calculation of ECC using \eqref{ecc} is a time-consuming procedure. It is observed from Fig. \ref{f3} that the results of the FDTD-CGF based techniques have excellent match with that obtained from far-field HFSS simulations. It should be noted that an approximated model of the dipoles consisting of some discrete current elements ($21$ in this case) is considered for the FDTD analysis. On the other hand, commercial FEM-based solver Ansys HFSS provides much more accurate representation of the system in terms of meshing. Yet it is observed that the FDTD-CGF based technique is capable of computing the ECC over wide frequency range with high resolution in considerably less amount of time ($\approx 120$ secs for the TD case). It can be inferred that although $\rho_{e,12}<0.1$ is achieved in the working band of the dipoles, the high degree of mutual coupling $S_{21}>-5$ dB renders the MIMO configuration of Fig. \ref{f1}(a) unsuitable for application.    
        
\subsection{Two Orthogonal Dipoles}
\begin{figure}[htbp]
\begin{center}
\includegraphics[width=13cm]{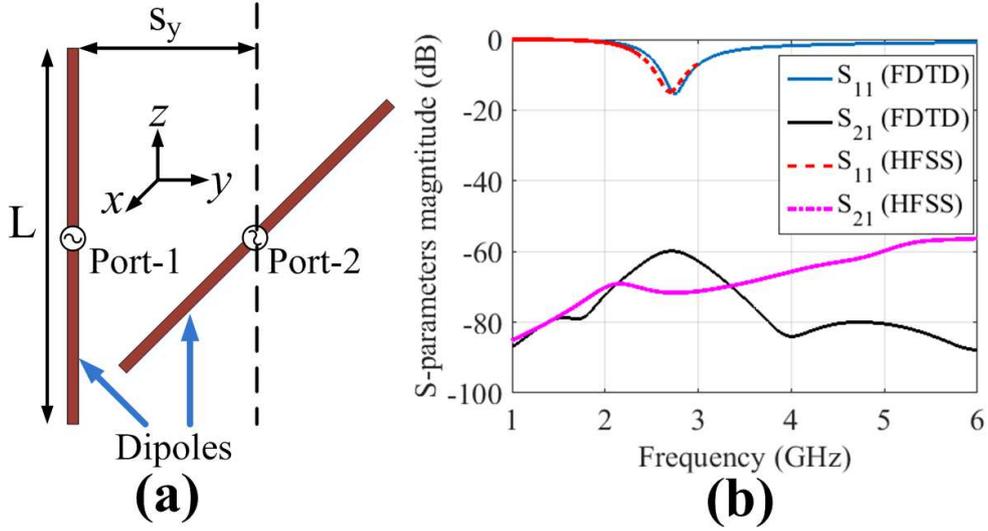}
\caption{\label{f4} (a) MIMO antenna system-2: Topology for the two identical thin-wire dipole antenna system arranged in orthogonal configuration in free-space. Here $L=50$ mm, $S_{y}=20$ mm. (b) Comparative plots of S-parameters (magnitude) with frequency, as obtained from FDTD and HFSS simulations.}
\end{center}
\vspace{-10pt}
\end{figure}

Fig. \ref{f4}(a) shows the second MIMO system under consideration which comprises of two half-wavelength dipoles, one oriented along $z$-direction, another along $x$ direction. Such orthogonal arrangements are useful for achieving omnidirectional pattern diversity as suggested in \cite{omni1}. The dimensions of the dipoles and inter-element spacing is kept as same as that for the parallel side-by-side scenario. The analysis of the dipole currents in time-domain is also carried out in a similar manner as described in previous subsection. For the $x$-directed dipole (dipole-2) the port voltage and time-domain current on the current elements are calculated by the following relations:

\begin{equation} \label{vin2}
V_{in,2}^{n}=-E_{x}(i_{cm},j_{cm},k_{cm})\Delta x
\end{equation}

\begin{multline} \label{Itot2}
I_{2}^{n+\frac{1}{2}}(i) = \Delta z \left( H_{z}^{n+\frac{1}{2}}(i,j_{cm}-\frac{1}{2},k_{cm})-H_{z}^{n+\frac{1}{2}}(i,j_{cm}+\frac{1}{2},k_{cm}) \right) \\ + \Delta y \left( H_{y}^{n+\frac{1}{2}}(i,j_{cm},k_{cm}+\frac{1}{2})-H_{y}^{n+\frac{1}{2}}(i,j_{cm},k_{cm}-\frac{1}{2}) \right)
\end{multline} 

\begin{figure}[htbp]
\begin{center}
\includegraphics[width=9cm]{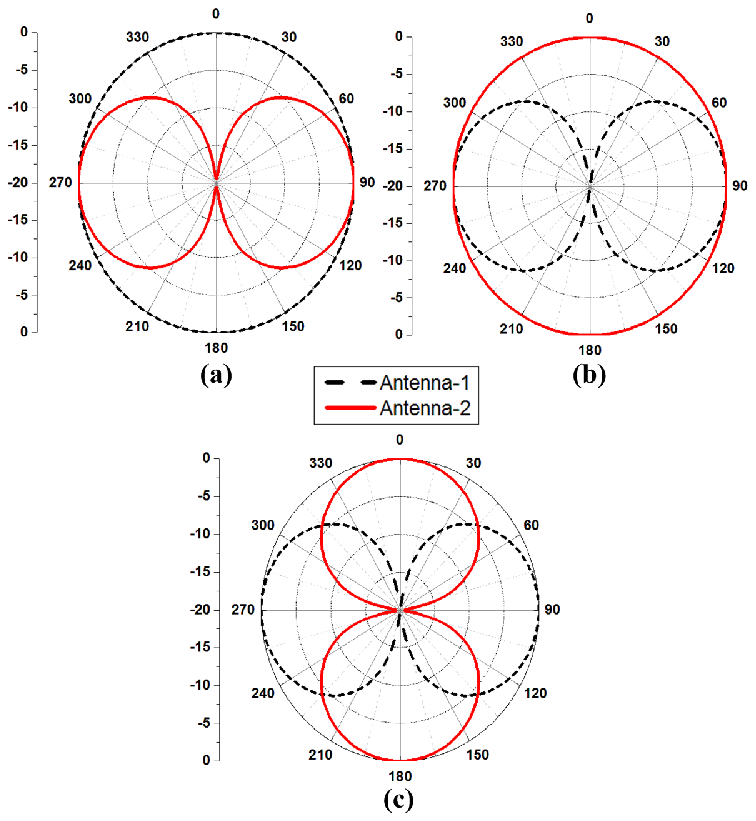}
\caption{\label{f5} Simulated total-gain patterns at resonance frequency, for both the antennas in the MIMO configuration of Fig. \ref{f4}(a) along the three principal planes: (a) $xy$-plane, (b) $yz$-plane and (c) $xz$-plane.}
\end{center}
\vspace{-10pt}
\end{figure} 

\begin{figure}[htbp]
\begin{center}
\includegraphics[width=9cm]{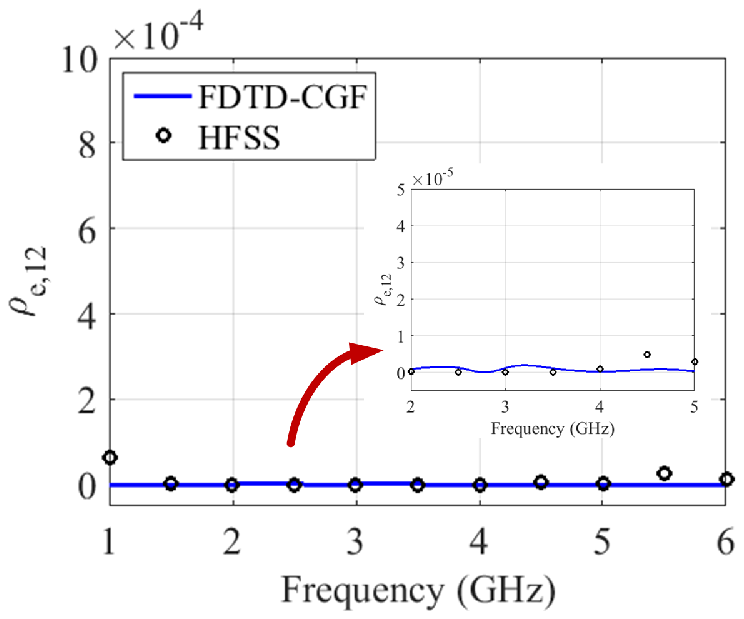}
\caption{\label{f6} Simulated variation of ECC for the MIMO configuration of Fig. \ref{f4}(a) over the frequency range $1-6$ GHz as obtained from FDTD-CGF techniques and Ansys HFSS. In Ansys HFSS, the ECC is calculated from far-field distribution at discrete frequency points.}
\end{center}
\vspace{-10pt}
\end{figure} 

Fig. \ref{f4}(b) depicts the magnitude of S-parameters for the MIMO system having orthogonal dipoles. Distinctive antenna resonance aroud $2.7$ GHz can be observed with very low mutual coupling ($S_{21}<-60$ dB) from both FDTD as well as HFSS simulation results. Fig. \ref{f5} shows the simulated radiation patterns of both the antennas along the three principal planes, calculated from the amplitude and phase of the antenna current densities at resonance frequency (Antenna-1 refers to the $z$-directed dipole and Antenna-2 refers to the $x$-directed dipole). The radiation pattern plots in Fig. \ref{f5} are also verified from Ansys HFSS. The plots Fig. \ref{f5}(a) to Fig. \ref{f5}(c) indicate good pattern omnidirectional diversity performance for the orthogonal MIMO configuration. The simulated ECC using both the FDTD-CGF techniques and separate HFSS simulations are shown in Fig. \ref{f6}. Both FDTD-CGF and HFSS simulations suggest that there is extremely low correlation ($\ll 10^{-3}$) between the two orthogonal dipole antennas, hence this is more applicable in a practical MIMO scenario as compared to the previously discussed side-by-side topology. 

\section{Comparison between FDTD-CGF-FD and FDTD-CGF-TD Techniques} \label{s4}
\begin{figure}[htbp]
\begin{center}
\includegraphics[width=13cm]{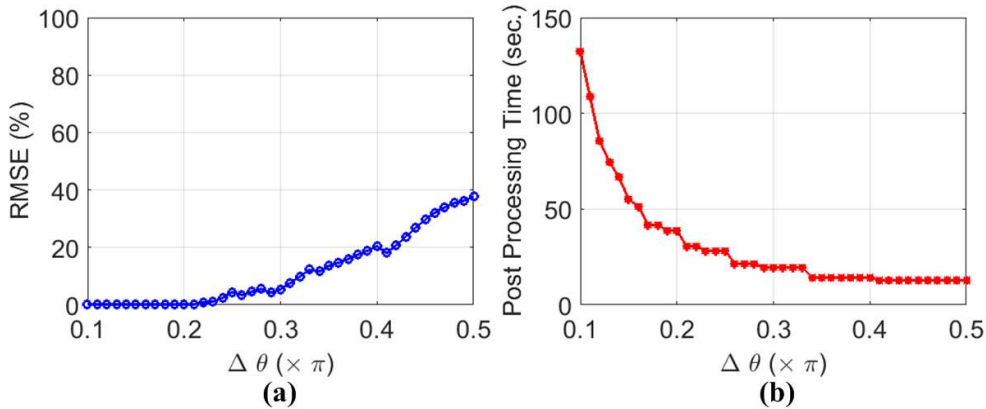}
\caption{\label{f7} (a) Variation of RMSE for FDTD-CGF-TD technique w.r.t FDTD-CGF-FD techinque, with angular steps $\Delta \theta (= \Delta \phi)$ used in the integrals of \eqref{psipqt} and \eqref{psimmpqt}. The MIMO configuration of Fig. \ref{f1}(a) is considered. (b) Plot of simulation time versus $\Delta \theta (= \Delta \phi)$ for FDTD-CGF-TD technique.}
\end{center}
\vspace{-10pt}
\end{figure}

It has been discussed in previous sections that for computing ECC over a wide range of frequencies with high degree of resolution, the FDTD-CGF-FD scheme takes much larger simulation time and the FDTD-CGF-TD is comparatively more efficient. To illustrate this point on simulation-time regarding wideband ECC calculation, both the FDTD-CGF-FD and FDTD-CGF-TD techniques are used for post-processing on the same time-domain data-set obtained from FDTD simulation of the MIMO system in Fig. \ref{f1}(a). The MATLAB code for post-processing is written a system having specifications: 
\begin{itemize}
\item Processor: Intel (R) Core (TM), i7-4770 CPU $@$ 3.40 GHz - 2.70 GHz
\item Installed RAM: 16.0 GB (15.7 GB Usable)
\item Operating system: 64 bit 
\end{itemize} 
The variation of ECC over a frequency range of 1 GHz to 6 GHz with frequency steps of 0.01 GHz is determined using both FDTD-CGF-FD and FDTD-CGF-TD techniques. It is observed that while time taken for FDTD-CGF-FD is $2831$ seconds, the FDTD-CGF-TD takes only $121$ seconds to obtain the same accuracy, which indicates almost $24$-times faster calculation in the TD post processing scheme. But in the FDTD-CGF-TD scheme, one important issue is the angular resolution ($\Delta \theta$ and $\Delta \phi$) needed for computation of the integrals in \eqref{psipqt} and \eqref{psimmpqt}. To justify our point, we calculate the root-mean-square-error (RMSE) in the FDTD-CGF-TD based results, considering FDTD-CGF-FD results as reference. Fig. \ref{f7} depicts how the percentage RMSE increases with lesser angular resolution, while the simulation time reduces (Fig. \ref{f7}(b)). But it is observed that RMSE around $0.17\%$ is achieved for an angular resolution of $\leq 0.1\pi$. 

\section{Conclusion} \label{s5}
In this paper, FDTD-CGF computation scheme is proposed by integrating FDTD with CGF calculation, for accurately determining the wideband frequency variation of ECC for MIMO antenna systems. Both frequency-domain (FD) and time-domain (TD) post-processing schemes are developed for post-processing in this proposed FDTD-CGF technique. It should be noted at this point that the concept of frequency domain solution of CGF using MoM was already introduced in [12], but the issue of wideband computation of ECC was not stressed upon there. In this work, CGF computation is done using currents obtained from FDTD, and a time-domain (TD) post-processing scheme is introduced along with the frequency-domain (FD) method. This is the first time, such TD post processing of CGF is reported. The FDTD-CGF-TD technique utilizes time-domain operations like time-reversal, convolution and integration to successfully reduce the computation time for wideband ECC calculation with high frequency resolution, as compared to the FDTD-CGF-FD method. It should be noted that the MIMO antenna topologies as well as FDTD simulation setup used in this work are much simplified, as they are mainly used to verify the correctness of the proposed FDTD-CGF-FD and FDTD-CGF-TD techniques. Improved FDTD schemes using adaptive meshing and better absorbing boundary conditions can enhance the applicability of these ECC computation techniques for more complex MIMO antenna configurations.    

\section*{Appendix-I: Expressions for Angular Cross-Correlation Component Functions}
The function $f_{pq}(\theta,\phi)$ where $p=x,y,z$ and $q=x,y,z$ can be expressed as:
$$f_{xx}(\theta,\phi)=\sin\theta (1-\sin^{2}\theta \cos^{2}\phi)$$
$$f_{yy}(\theta,\phi)=\sin\theta (1-\sin^{2}\theta \sin^{2}\phi)$$
$$f_{zz}(\theta,\phi)=\sin\theta (1-\cos^{2}\theta) $$
$$f_{xy}(\theta,\phi)=f_{yx}(\theta,\phi)= -\sin^{3}\theta \cos\phi \sin\phi$$
$$f_{xz}(\theta,\phi)=f_{zx}(\theta,\phi)= -\sin^{2}\theta \cos\theta \cos\phi$$
$$f_{yz}(\theta,\phi)=f_{zy}(\theta,\phi)= -\sin^{2}\theta \cos\theta \sin\phi$$

\section*{Acknowledgment}
The authors are extremely grateful to Prof. Y. M. M. Antar, Department of Electrical and Computer Engineering Royal Military College of Canada, for motivating us to work in this domain, during his visit to IIT Kanpur under the IEEE AP-S Distinguished Lecture Program.


\begin{thebibliography}{99}
\bibitem{mimo1} L. Swindlehurst, E. Ayanoglu, P. Heydari, and F. Capolino, ``Millimeter-Wave Massive MIMO: The Next Wireless Revolution?,'' \textit{IEEE Communications Magazine}, pp. 56-62, September 2014.


\bibitem{mimo2} T. S. Rappaport, S. Sun, R. Mayzus, , H. Zhao, Y. Azari, K. Wang, G. N. Wong, J. K. Schulz, M. Samimi and F. Gutirrezi, ``Millimeter Wave Mobile Communications for 5G Cellular: It Will Work!,'' \textit{IEEE Access}, Vol. 1, pp. 335-349, 2013.

\bibitem{mimo3} M. S. Sharawi, ``Printed Multi-Band MIMO Antenna Systems and Their Performance Metrics,'' \textit{IEEE Antennas and Propagation Magazine}, Vol. 55, pp. 218-232, 2013. 

\bibitem{mimo4} M. A. Jensen and J. W. Wallace, ``A Review of Antennas and Propagation for MIMO Wireless Communications,'' \textit{IEEE Transactions on Antennas and Propagation}, Vol. 52, No. 11,  pp. 2810-2824, 2004.
 
\bibitem{mimo5} E. Rajo-Iglesias, O. Quevedo-Teruel and M. SanchezFernandez, ``Compact Multi-Mode Antenna for MIMO Applications,'' \textit{IEEE Antennas and Propagation Magazine}, Vol. 50, No. 2,  pp. 197-205, 2008. 

\bibitem{mimo6} O. Quevedo-Teruel, M. Sanchez-Fernandez, M. L. PabloGonzalez and E. Rajo-Iglesias, ``Alternating Radiation Patterns to Overcome Angle of Arrival Uncertainty,'' \textit{IEEE Antennas and Propagation Magazine}, Vol. 52, No. 1, pp. 70-79, 2010.

\bibitem{ecc1} M. P. Karaboikis, V. C. Papamichael, G. F. Tsachtsiris, C. F. Soras and V. T. Makios, ``Integrating Compact Printed Antennas Onto Small Diversity/MIMO Terminals,'' \textit{IEEE Transactions on Antennas and Propagation}, Vol. 56, No. 7, pp. 2067-2078, 2008.

\bibitem{ecc2} S. Blanch, J. Romeu, and I. Corbella, ``Exact representation of antenna system diversity performance from input parameter description,'' \textit{Electronics Letters}, vol. 39, no. 9, pp. 705–707, May 2003.

\bibitem{ecc3} P. Hallbjorner, ``The significance of efficiencies when using S-parameters to calculate received signal correlation from two antennas,'' \textit{IEEE Antennas Wireless Propagation Letters}, vol. 4, pp. 97–99, Jun. 2005.

\bibitem{ecc4} M. S. Sharawi, ``Printed MIMO antenna systems: Performance metrics, implementations and challenges,'' \textit{Forum for Electromagn. Res. Methods Appl. Technol. (FERMAT)}, vol. 1, 2014 [Online]. Available: http://www.e-fermat.org/

\bibitem{ecc5} T. Wittig and V. Sokol, ``MIMO Antenna Simulation,'' UGM 2009, Darmstadt, Germany, 2009 [Online]. Available: https://www.cst.com/Content/Events/UGM2009/3-1-4-MIMO-Antenna-Simulation.pdf

\bibitem{antar1} S. M. Mikki and Y. M. M. Antar, ``On Cross Correlation in Antenna Arrays  With Applications to Spatial Diversity and MIMO Systems,'' \textit{IEEE Transactions  on Antennas and Propagation}, Vol. 63, No. 4, pp. 1798-1810, 2015. 

\bibitem{antar2} S. Clauzier, S. M. Mikki, and Y. M. M. Antar, ``A Generalized Methodology for Obtaining Antenna Array Surface Current Distributions With Optimum Cross-Correlation Performance for MIMO and Spatial Diversity Applications,'' \textit{IEEE Antennas and Wireless Propagation Letters}, Vol. 14, pp. 1451-1454, 2015.
 
\bibitem{antar3} S. Clauzier, S. M. Mikki and Y. M. M. Antar, ``Design of high-diversity gain MIMO antenna arrays through surface current optimization,'' \textit{Proceedings of IEEE International Symposium on Antennas and Propagation and USNC/URSI National Radio Science Meeting}, pp. 9-10, 2015. 

\bibitem{antar4} S. Clauzier, S. M. Mikki and Y. M. M. Antar, ``Generalized methodology for antenna design through optimal infinitesimal dipole model,'' \textit{Proceedings of International Conference on Electromagnetics in Advanced Applications (ICEAA)}, pp. 1264-1267, 2015.

\bibitem{mant1} Q. Li, A. P. Feresidis, M. Mavridou, P. S. Hall, ``Miniaturized Double-Layer EBG Structures for Broadband Mutual Coupling Reduction
Between UWB Monopoles,'' \textit{IEEE Transactions on Antennas and Propagation}, Vol. 63, No. 3, pp. 1168-1171, 2015.

\bibitem{mant2} W. Liao, C. Hsieh, B. Dai, B. Hsiao, ``Inverted-F/Slot Integrated Dual-Band Four-Antenna System for WLAN Access Points,'' \textit{IEEE Antennas and Wireless Propagation Letters}, Vol. 14, pp. 847-850, 2015.

\bibitem{mant3} G. Zhai, Z. N. Chen and X. Qing, ``Enhanced Isolation of a Closely Spaced Four-Element MIMO Antenna System Using Metamaterial
Mushroom,'' \textit{IEEE Transactions on Antennas and Propagation}, Vol. 63, No. 8, pp. 3362-3370, 2015.

\bibitem{fdtd1} K. Ouchida, N. Honma and Y. Tsunekawa, ``Fast FDTD analysis of MIMO channel using spread spectrum technique,'' \textit{Proceedings of Asia-Pacific Microwave Conference (APMC)}, pp. 389-391, 2013.

\bibitem{fdtd2} J. Zhao, J. Zhou and X. Yin, ``Multi-polarized MIMO indoor channel analysis based on 3-D CE-ADI-FDTD method,'' \textit{Proceedings of IEEE Asia-Pacific Conference on Antennas and Propagation (APCAP)}, pp. 5-6, 2012.

\bibitem{fdtd22} M. A. Jensen and Y. Rahmat-Samii, ``Performance Analysis of Antennas for Hand-Held Transceivers Using FDTD,'' \textit{IEEE Transactions on Antennas and Propagation}, vol. 42, no. 8, 1994. 

\bibitem{dgf1} J. Vazquez and C. G. Parini, ``Discrete green’s function formulation of FDTD method for electromagnetic modelling,'' \textit{Electronics Letters}, vol. 35, no. 7, pp. 554-555, 1999. 

\bibitem{dgf2} W. Ma, M. R. Rayner, and C. G. Parini, ``Discrete green’s function formulation of the FDTD method and its application in antenna modeling,'' \textit{IEEE Transactions on Antennas and Propagation}, vol. 53, no. 1, pp. 339-346, 2005. 

\bibitem{dgf3} T. P. Stefanski, ``Hybrid technique combining the FDTD method and its convolution formulation based on the discrete green’s function,'' \textit{IEEE Antennas and Wireless Propagation Letters}, vol. 12, pp. 1448-1451, 2013.

\bibitem{dgf4} S. Mirhadi, M. Soleimani, and A. Abdolali, ``UWB antennas analysis using FDTD-based discrete green’s function approach,'' \textit{IEEE Antennas and Wireless Propagation Letters}, vol. 12, pp. 1089-1093, 2013.

\bibitem{dgf5} T. P. Stefanski, ``A New Expression for the 3-D Dyadic FDTD-Compatible Green’s Function Based on Multidimensional  Z-Transform,'' \textit{IEEE Antennas and Wireless Propagation Letters}, Vol. 14, pp. 1002-1005, 2015. 

\bibitem{fourier} B. P. Lathi, Linear Systems and Signals, Berkeley Cambridge Press, 2nd Edition.


\bibitem{fdtd3} D. M. Sullivan, Electromagnetic Simulation Using the FDTD Method, IEEE Press, New York, 2000.

\bibitem{fdtdds} D. Sarkar, S. Sahu, R. Ghatak, R. K. Mishra and D. R. Poddar, ``FDTD Analysis of Coupled Microstrip Lines Separated by a DNG Slab,'' \textit{Proceedings of Loughborough Antenna and Propagation Conference (LAPC-IEEE)}, pp. 398-400, 2010.
 
\bibitem{pozar} D. M. Pozar, Microwave Engineering, John Wiley, Fourth Edition, 2011.  

\bibitem{fdtd4} B. E. Bharatha Yajaman, FDTD Modeling Of RF And Microwave Circuits With Active And Lumped Components, Masters Thesis, Texas Tech University, 2004.

\bibitem{omni1} J. Malik, D. Nagpal and M.V. Kartikeyan, ``MIMO antenna with omnidirectional pattern diversity,'' \textit{Electronics Letters}, Vol. 52, No. 2, pp. 102-104, 2016.  


\end{thebibliography}
\end{document}